\documentclass[aps,prl,twocolumn,superscriptaddress,showpacs,showkeys,floatfix]{revtex4-1}
\usepackage{epsfig}
\usepackage{graphicx}
\usepackage{amsmath}
\usepackage[english]{babel}
\usepackage{multirow}
\usepackage{placeins}

\usepackage[dvips]{color}

\begin{document}

\title{Characterizing departure delays of flights in passenger aviation network of United States}

%% \title{Facilitation and Inhibition of Network Percolation by Distance-Dependent Strategy in Two Dimensions}

\author{Yan-Jun Wang\dag *}
\affiliation{College of Civil Aviation, Nanjing University of Aeronautics
and Astronautics, Nanjing, 210016, China}
\affiliation{College of Science, Nanjing University of Aeronautics
and Astronautics, Nanjing, 210016, China}

\author{Ya-Kun Cao\dag}
\affiliation{College of Science, Nanjing University of Aeronautics
and Astronautics, Nanjing, 210016, China}

\author{Chen-Ping Zhu\*}
\affiliation{College of Science, Nanjing University of Aeronautics
and Astronautics, Nanjing, 210016, China}

\author{Fan Wu}
\affiliation{College of Civil Aviation, Nanjing University of Aeronautics
and Astronautics, Nanjing, 210016, China}
\affiliation{Air Traffic Management Bureau of Northwest China, Xi'an, 710082, China}

\author{Ming-Hua Hu}
\affiliation{College of Civil Aviation, Nanjing University of Aeronautics
and Astronautics, Nanjing, 210016, China}

\author{Baruch Barzel}
\affiliation{Mathematics Department, Bar Ilan University, Lamat Gan, Israel}
%\affiliation{Department of Physics, Boston University,  United States of America}
%\affil[*]{oldpigman1234@126.com}
%\affil[*]{yanjunwang@gmail.com}
%\affil[*]{hes@bu.edu}
\author{H. E. Stanley\*}
\affiliation{Center for Polymer Studies and Department of Physics, Boston University, Boston, Massachusetts 02215, USA
}

\date{\today}

\begin{abstract}
Flight delay happens every day in airports all over the world. However, systemic investigation in large scales remains a challenge. We collect primary data of domestic departure records from Bureau of Transportation Statistics of United States, and do empirical statistics with them in form of complementary cumulative distributions functions (CCDFs) and transmission function  of the delays. Fourteen main airlines are characterized by two types of CCDFs: shifted power-law and exponentially truncated shifted power-law. By setting up two phenomenological models based on mean-field approximation in temporal regime, we convert effect from other delay factors into a propagation one. Three parameters meaningful in measuring airlines emerge as universal metrics. Moreover, method used here could become a novel approach to revealing practical meanings hidden in temporal big data in wide fields.
\end{abstract}


\maketitle

Flight delays happen every day in the airports all over the world. Passengers in different countries suffer from such ubiquitous events, and economy and traffics of many cities are largely harmed by these events. Delays of individual flights seem to be random at a glance. Actually, they obey certain statistical laws\cite{Fleurquin}  taking a long-term delay records for a large number of flights into account. In general, some delay originating from an upstream flight spreads to downstream flights, which is particularly evident when an aircraft continues its tasks of successive flights. This phenomenon is defined as delay propagation(DP) \cite{Kafle16,Ahmad08,Ahmad10,Wang03}. To alleviate the effect caused by DP \cite{Ahmad08,Ahmad10,Lan}, we need to characterize them quantitatively first, which is still a big difficult task now although empirical statistics \cite{Fleurquin,Ahmad08,Ahmad10,Beatty,Campanelli,Fleurquin13,Liu} and mechanism modeling \cite{Kafle16,Wang03,Fricke,Abdelghany,Pyrgiotis,Tu,Wong} have appeared recently. The key problems remaining challenging us lie on: what kind of statistical law we can learn from the large scale historical records of flight-delay events? And what are possible mechanisms underlying them in terms of propagation and other factors? They motivate our present work from both aspects of empirical statistics and theoretical modeling.

The causes of flight delays have been classified into five categories by the US Bureau of Transportation Statistics (BTS) \cite{BTS}: (1)Air carrier; (2)extreme weather; (3)national aviation system(NAS)referring heavy traffic volume and ait traffic control; (4)security; (5)late arriving aircraft. Just as cited by Baumgarten et al. \cite{Baumgarten}, the propagation factor(PF)(delay propagation caused by late arrival of last flight of the same aircraft) occupies 35 percent in this classification. In sense of propagation, we can roughly divide them into just two categories: PF and non-propagation factor(NPF). %which is the fifth one in the classification of BTS.
%Here we take the concept of DP in its narrow sense, instead of which meaning a flight's delay spreading to other flights even over the whole network.

The statistics on empirical data of flight delays makes up the primary step on the characterization of DP \cite{Fleurquin,Ahmad10,Beatty,Campanelli,Fleurquin13}. Analytical model \cite{Kafle16,Wang03,Fricke,Abdelghany,Pyrgiotis,Tu,Wong} serves as the second step on the investigation of DP. In view of successful contribution from these two steps, we are called to go on the further step, i.e., to make general models to explore generic mechanisms based on airline-specific big data of flight delays.

We start to investigate empirical laws from collecting data of departure delays (DD). Domestic DD data in all the airports over the whole United States are downloaded from BTS of America \cite{BTS}, including flight numbers, tail numbers, scheduled departure and arrival time, and actual departure and arrival time of each flight in the American passenger network. And DD in 2014 of each flight is calculated as the difference between actual departure time and the scheduled one. Then we obtain the statistical results of probability distribution functions (pdf) of DD time for each airline, respectively. Fig.1 in $\bf{Supplementary Information(SI)}$ illustrates the pdfs for 14 main airlines of America. Furthermore, integrating pdf from certain DD to infinity makes up complementary cumulative distribution functions (CCDFs) \cite{Kwak} for every airline. These CCDFs are shown with color filled circles in Fig.1 and Fig.2, respectively. Each of them is set up a model to understand the underlying mechanism.

%%%%%%%%%%%%%%%%%%%%%%%%%%%%%%%%%%%%%%%%%%%%%%%%%%%%%%%%%%%%%%%%%%%%%%

\section*{Results}

\subsection*{1. Model 1}

As a rough approximation, we start to describe DD of flights in a unified way, i.e., to account the effect from NPF by converting them into that equivalent to PF. Then, we merely consider the effect of PF in the derivation of CCDFs of airlines. At first, we make two assumptions based on data observation from the spirit of mean field approach. As assumption (1), current DD caused by PF is assumed smaller than last DD of the same aircraft in general cases over one-year statistics, which takes account of the DP and the efforts against it made by the staffs of the management and operation. In practical manipulation, we take DP as follows: for a flight with DD $\it{l}$ (in minutes), if last flight operated by the same aircraft is larger than $\it{l}$  we assume that the current DD is caused by PF. This is because real delays incurred by PF are not discernable at present\cite{Abdelghany}. Based on such an assumption, we let the number of flights with DP in the unit delay interval near $\it{l}$ be divided by the total number of the flights with the delay larger than $\it{l}$, and define this ratio as the probability of delay transmission $q(\it{l})$. Then
\begin{equation}
n_{B}(\it{l}) = q(\it{l}) \int_{\it{l}}^{\infty} n(\it{l})d\it{l}
\end{equation}
where $n_{B}(\it{l})$ is the number of delayed flights in the unit interval near $\it{l}$ caused by DP, and $n(\it{l})$ is the total number of delayed flights per delay interval near $\it{l}$, therefore, cumulated number of delayed flights
$N(\it{l}) = \int_{\it{l}}^{\infty} n(\it{l})d\it{l}$. By formula (1) we account the number of propagation-caused delayed flights in the unit interval of $(\it{l}$ as the transmission result of all delayed flights with from $\it{l}$ to $\infty$.
In the present phenomenological model, the form of the function $q(\it{l})$ is unknown and needs to be dug out by data-mining later. As a simple result of direct observation, the probability of large delay obviously decrease with $\it{l}$  monotonously, so the probability $q(\it{l})$ of delay transmission per delay interval should be a monotonous decreasing function as shown in Fig.2 of $\bf{SI}$, for instance of the real data of airline AA. However, it should be finite in the limit $\it{l} \rightarrow 0$. Therefore, it is a reasonable choice to set
\begin{equation}
q(\it{l}) = \frac{\alpha}{\it{l}+\beta}
\end{equation}
as its tentative form, with $\alpha$, $\beta$ as airline - specific parameters determined by data-mining according to formula (2). In our phenomenological model, the parameters in $q(\it{l})$ have to be obtained from the data. Taking the American Airlines (AA) as an example, we calculate the value of $q(\it{l})$ for 237 aircrafts in the whole year of 2014 using formula (2). By taking $\Delta \it{l}= 10 $ min and select (0,10) (10,20) ..., as statistical interval, then letting the variable $\it{l}$ be equal to the median of each interval, we plot empirical results in Fig.2 of $\bf{SI}$.

As assumption (2), we suppose that propagation-caused fraction of delayed flights per delay interval keeps a constant $\it{k}$, so that we have
\begin{equation}
n_B(\it{l}) = \it{k} * n(\it{l})
\end{equation}
Comparing formula (1) with (3), we see
\begin{equation}
 n_B(\it{l}) = \it{k} n(\it{l}) = q(\it{l})N(\it{l})
\end{equation}
that is, $\it{k}*(N(\it{l})- N(\it{l}+ d\it{l}))/d\it{l} = q(\it{l})N(\it{l})$ , we have

\begin{equation}
\frac{dN(\it{l})}{d\it{l}} = \frac{q(\it{l})}{- \it{k}} N(\it{l})
\end{equation}

When the tentative transmission function $q(\it{l})$ is substituted into it, we obtain
\begin{equation}
\frac{dN(\it{l})}{d\it{l}} = \frac{\alpha}{- \it{k}(\it{l}+\beta)} N(\it{l})
\end{equation}
Therefore,
\begin{equation}
 N(\it{l}) = c (\it{l}+\beta)^{- \alpha_0}
\end{equation}
where  $\alpha_0 = \alpha/\it{k}$. And the CCDF of flight DD reads
\begin{equation}
P(\it{l}_0 > \it{l}) = \frac{N(\it{l})}{N_0} = c_0 (\it{l}+\beta)^{-\alpha_0}
\end{equation}
where $N_0$ is the total number of flights of an airline over a year. Therefore, we obtain shifted power-law (SPL)\cite{Chang,Fu,Wang} to describe the CCDFs of DD of flights.

Formula (8) in phenomenological Model 1 is checked for 14 airlines(See Fig.1 of $\bf{SI}$). It is valid for AA, MQ, F9, DL, HA and AS airlines with each pair of two parameters $\beta_1$ and $\alpha_0$ well adjusted to fit the primary data from a specific airline, where $\beta_1$ means adaptive $\beta$  and $\alpha_0$ in formula (8)(see Fig.1). However, it fails to cover the remaining 8 airlines, which demands other types of mechanisms. In the panels of Fig.1, real CCDFs of these 6 airlines are plotted with the filled circles in black, green, blue, wine red, dark yellow and purple, red lines representing analytic results in formula (8) with corresponding shift parameters $\beta_1 = 40, 50, 90, 40, 30$ and $90$ fit linear parts of empirical data well. While the transmission function $q(\it{l})$ is checked in its integrated form $F(\it{l})$ following formula (2) with real data, where $\beta_2$ means $\beta$ in formula (8) adaptive to fitting integrated $q(\it{l})$ ($F(\it{l}) = \int_{0}^{\it{l}} q(\it{l})d\it{l}$) in Fig.3(also collected in Table (1) of $\bf{SI}$). The tail-derivations of red lines from filled circles in both Fig.1 and Fig.3 are discussed in $\bf{SI}$. In fitting these CCDFs from primary data with the present mechanism of DP, we are actually converting the effect from NPF into an equivalent one from PF, since the CCDFs illustrated by colored symbols contain effect from all kinds of delay factors, while the red lines are based on the all-PF assumption potentially.
%\usepackage{graphicx}

%\begin{center}
%\begin{figure}
%[htbp!]
%\centering
%\subfloat[Airport Movements in ZSNJ airport during 10:00-12:00]{\label{fig:zsnjmov}\includegraphics[width=0.5 \textwidth]{fig/ZSNJ_MOVE_SIM}}
%\subfloat[Departure and arrival delay in ZSNJ airport during 10:00-12:00]{\label{fig:zsnjdelay}
%\includegraphics[width=0.8 \textwidth]{Fig3.EPS}
% \\
%\subfloat[Flights delay during 10:00-12:00 (without ATFM constraints)]{\label{fig:zsnjno}\includegraphics[width=0.5 \textwidth]{fig/12-16}}
%\subfloat[Flights delay during 12:00-14:00 (after ATFM contraints)]{\label{fig:zsnjatfm}\includegraphics[width=0.5 \textwidth]{fig/no12-16}}
%\caption{Comparisons on delay in the national airports after the implementation of ATFM}
%\label{Fig.1}
%\end{figure}
%\end{center}

\begin{figure*}
\centering
%\scalebox{0.4}[0.4]{\includegraphics[trim=0 0 0
%0,clip]{Fig3.EPS}}
\includegraphics[width=0.9\linewidth]{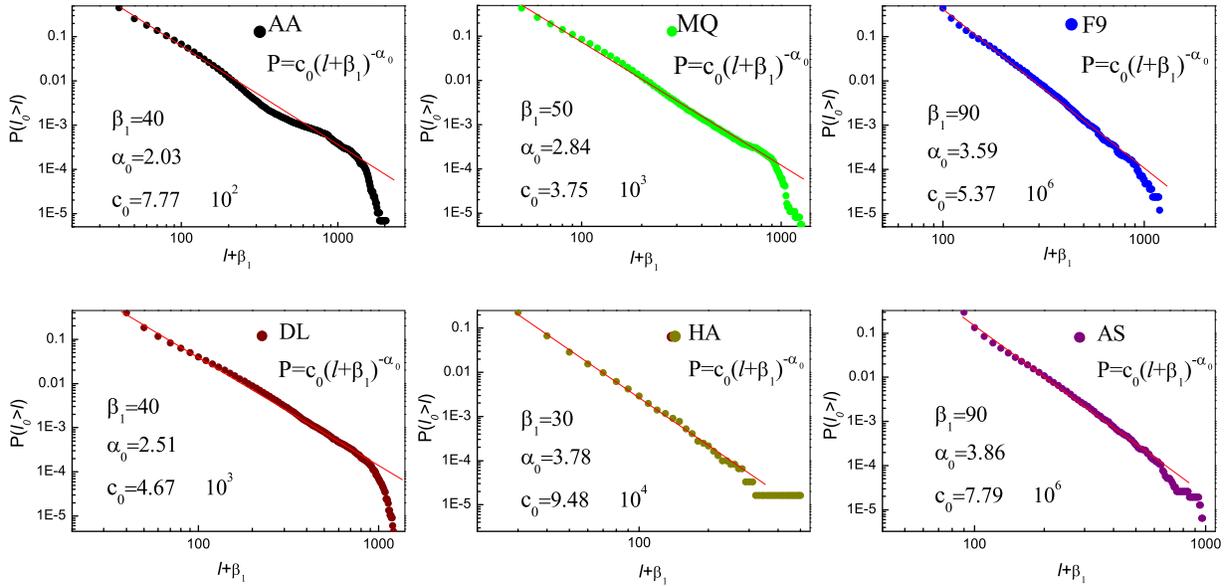}
\caption{
Log-Log plots of CCDFs of 6 airlines in US.
Empirical data decay with delay $\it{l}$ plus shift parameter $\beta_1$
linearly in the main body for the region with small $\it{l}$, where $\beta_1$ means fitting CCDFs using $\beta$ in formula (8). Filled circles in black, green, blue, wine red, dark yellow and purple illustrate CCDFs integrated pdfs from real data in Fig.1 of $\bf{SI}$ for airlines AA, MQ, F9, DL, HA and AS, respectively. Red lines illustrate the CCDFs in formula (8) of Model 1. By adapting parameters $(\alpha_0, \beta_1)$ in Model 1 with the mechanism of DP, we fit the empirical data. The shift parameters are $\beta_1 = 40, 50, 90, 40, 30,$ and $90$ for airlines AA, MQ, F9, DL, HA and AS, respectively. $\Delta \it{l} = 10$ $min$.}
\end{figure*}

\begin{figure*}[ht]
\centering
\includegraphics[width=0.9\linewidth]{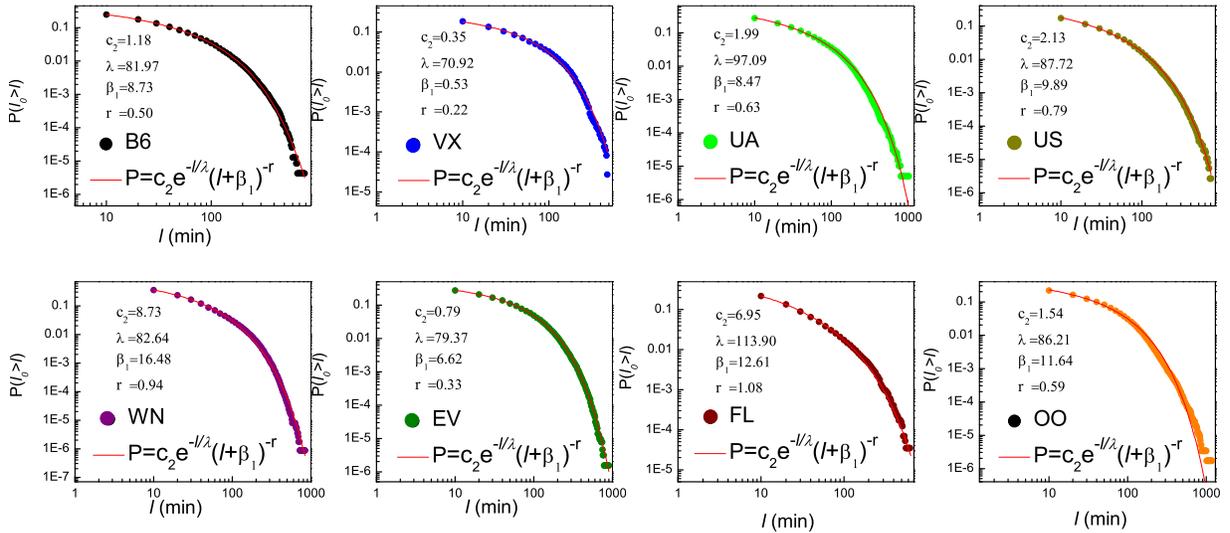}
\caption{Log-Log plots of CCDFs of airlines in US. Empirical data decay with delay $\it{l}$ plus shift parameter $\beta_1$ monotonically in truncated exponential function $e^{- \it{l}/\lambda}$, where $\beta_1$ means fitting CCDFs of Fig.2 using $\beta$ in formula (14). Filled circles in black, blue, green, dark yellow, purple, olive, wine red and orange illustrate CCDFs from real data (Fig.1 of $\bf{SI}$) of airlines B6, VX, UA, US, WN, EV, FL and OO, respectively. Red lines illustrated the CCDFs in formula (14). By adapting parameters $(\lambda, \beta_1$ and $r)$ in Model 2 with modified mechanism of DP, red lines from formula (14) fit empirical data with
$\lambda = 81.97, 10.92, 97.09, 87.72, 82.64, 79.37, 113.90$ and $86.21$ for these 8 airlines, respectively. $\Delta \it{l} = 10$ $min$.
}
\label{fig:5}
\end{figure*}

\begin{figure*}
\centering
%\scalebox{0.4}[0.4]{\includegraphics[trim=0 0 0 0,clip]{Fig4.eps}}
\includegraphics[width=0.9\linewidth]{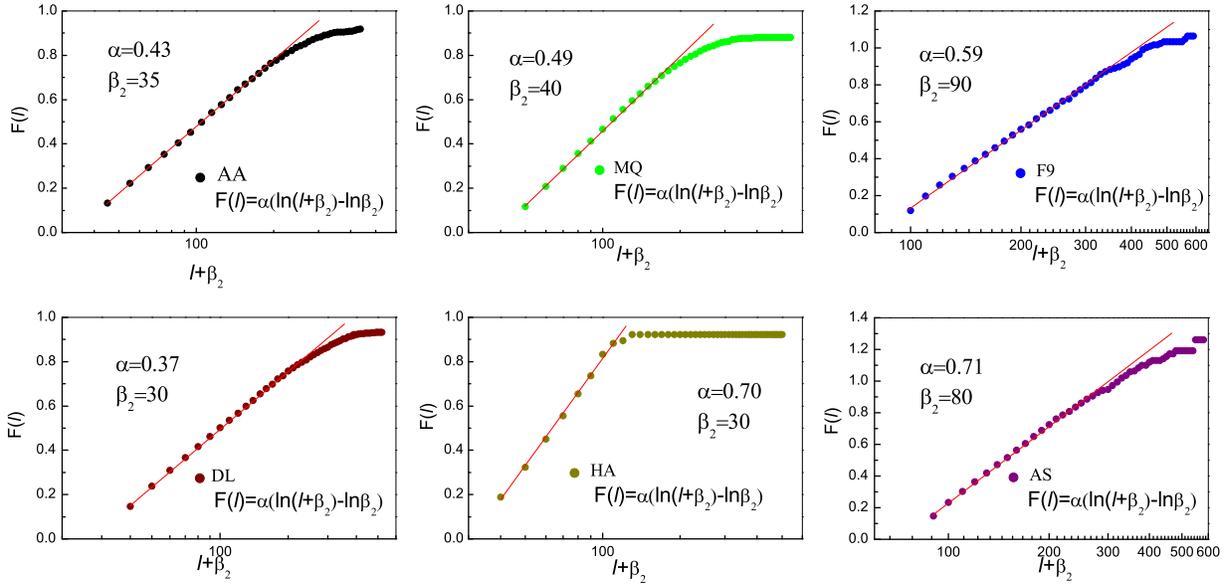}
\caption{Log-Linear plots of the functions $F(\it{l})$ which represent
the integrated form of transmission function $q(\it{l})$ in formula (2). Color symbols in black, blue, green, dark yellow, purple, olive, wine red and orange illustrate the results directly cumulated from empirical data based on the definition (formula (2)) for airlines AA, MQ, F9, DL, HA and AS, respectively. While red lines are results integrated from $q(\it{l})$ in formula (2) with their parameters $\beta_2$ approaching parameter $\beta_1$ in Fig.1 correspondingly. $\Delta \it{l} = 10$ $min$.}
\end{figure*}

\begin{figure*}[ht]
\centering
\includegraphics[width=0.9\linewidth]{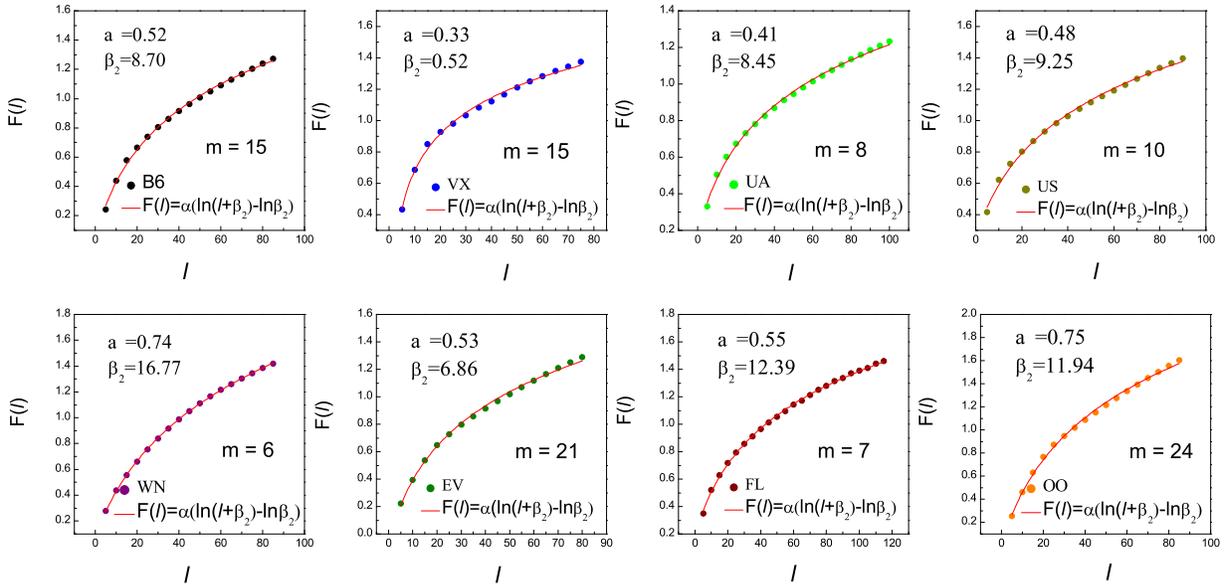}
\caption{Functions $F(\it{l})$ represent cumulated form of transmission probability $q(\it{l})$  based on formula (2) with compensation interval $m$ shown for each airline. Color symbols illustrate the results below $\lambda$ directly cumulated from real data based on the definition but with specific compensation interval $m$ for airlines B6, VX, UA, US, WN, EV, FL and OO, respectively. While red lines are results integrated from formula (2) with their parameters $\beta_2$ approaching the parameters $\beta_1$ in Fig.2 correspondingly.
$\Delta \it{l} = 5$ $min$.
}
\label{fig:6}
\end{figure*}

\subsection*{2. Model 2}

     To understand the CCDF data from the remaining 8 airlines, we need a new mechanism other than what is for those 6 ones. At first, the assumption of a constant $\it{k}$ for the propagation-caused fraction of delayed flights per delay interval in model 1 can not be valid for complex cases. Actually, formula (5) would be still valid if $\it{k}$ behaves in an infinitesimal in the same order of $q(\it{l})$ as $\it{l} \rightarrow \infty $ according to $L'Hospita$ rule. Therefore, we can expect that $\it{k}$ has a similar form of $ q(\it{l})$. Tentatively, we let

\begin{equation}
\it{k}(\it{l}) = \frac{1}{g \it{l}+h},
\end{equation}
where $g$ and $h$ are both phenomenological constants. Substituting it into formula (5), we obtain
\begin{equation}
\frac{dN(\it{l})}{d\it{l}}
= \frac{-\alpha(g\it{l}+h)}{\it{l}+\beta} N(\it{l})
\end{equation}
then
\begin{equation}
\frac{dN(\it{l})}{N(\it{l})}
= (-g\alpha + \frac{-r}{\it{l}+\beta}) d\it{l}
\end{equation}
where, $r=\alpha(h-g\beta)$. We have
\begin{equation}
ln N(\it{l}) = -g\alpha\it{l}-rln(\it{l}+\beta)+c_0
= ln e^{-g\alpha \it{l}}+ln(\it{l}+\beta)^{-r} + c_0
\end{equation}
where $c_0$ is a constant. Therefore,
\begin{equation}
N(\it{l}) = c_1 e^{-\frac{\it{l}}{\lambda}} (\it{l}+\beta)^{-r}
\end{equation}
where $\lambda =1/g\alpha$, $c_1 = e^{c_0}$ .
And CCDF of DD of flights reads
\begin{equation}
P(\it{l}_0 > \it{l})=\frac{N(\it{l})}{ N_0 } =c_2 e^{-\frac{\it{l}}{\lambda}}(\it{l}+\beta)^{-r}
\end{equation}
where  $c_2 =c_1/N_0$.
Different from SPL in formula (8), in this mechanism we have a CCDF in the form of exponentially truncated shifted power-law (ETSPL) \cite{Gonzalez} for the remaining 8 airlines. When we fit empirical results with parameters in analytic functions based on DP, we convert effect by NPF into equivalent effect by DP factor just as in the way of Model 1.

In last model, we assumed that last DD of the same aircraft is greater than current one . In reality, however, the effects of NPFs including cancellations may overlap with PF. At present, we are not able to separate two kinds of factors in the empirical data\cite{Abdelghany}. Therefore, the empirical statistical results for 8 airlines in Fig.2 definitely contain effect from all kinds of delaying factors, including that of last delay smaller than the current one of the same aircraft. Obviously, assumption (1) in last model leads to a smaller value of $n_{B}(\it{l})$ than the actual one. Based on this consideration, we modify the definition of $n_{B}(\it{l})$ in Model 1:  the total number of the flights with the delay near $\it{l}$ per delay interval, but transmitted by last flight's DD greater than $(\it{l} - m)$, where $m$ describes average additional DD affecting current flights. By counting larger interval as effective one affecting the current delay of flights from equivalent DP, we compensate the deficient counting of contribution from all factors. We demand $\beta_2$ approaching values of $\beta_1$ in Fig.2, where $\beta_2$ means $\beta$ in formula (14) adaptive to fitting $F(\it{l})$ in Fig.4. Airline-specific parameters $m$ emerge out as the new metrics £¨see Fig.4).

 %due to effect from all kinds of NPFs and PF, which is an airline-specific parameter characterizing their operation features. Based on the parameters $\beta_1$ got from Fig.2 and affiliating different $m$ in the statistics of $q(\it{l})$ with newly defined $N_B(\it{l})$ over $N(\it{l})$ per statistical interval, we obtain the functions hence their integrated forms in Fig.4 with the symbols of filled circles in colors black, blue, green, dark yellow, purple, olive, wine red, and orange for airlines  B6, VX, UA, US, WN, EV, FL and OO, respectively. Then, we fit the $q(\it{l})$ (formula (12) with the parameters $g$ and $h$, respectively) to the empirical data calibrated by the new starting points $(\it{l} - m)$ of statistics in both their integrated forms(see Fig.4),where $F(\it{l}) = \int_{\it{l}}^{\infty} q(\it{l}) d\it{l}$ and get the compensation interval $m$ for each of 8 airlines shown in Fig.4 as well as Table.2 of SI. Virtually, here we convert the effect from all kinds of delay factors including all possible interactions among different flights in sectors and airports into an effective DP mechanism featured by the compensation interval $m$, which can be accounted as a mean-field approach in the temporal regime.

\subsection*{3. Aviation meanings of key parameters}

 Shift parameters $\beta_1$ in fitting CCDFs can serve as the feature quantities in minutes characterizing specific airline with the aviation meanings. We define the average flight delay absorption\cite{absorption1,absorption2} time $L$ (in minutes)(See Table (2) of $\bf{SI}$) of propagation delay as the average differences between last delay and the current one of the same aircraft for all flights with larger last delays in each airline. From direct statistics on empirical data, we find that the relation for the absorption time $L$ versus $\beta_1$ is a negatively correlated linear function. One can see from Fig.5(a) that the relation $L(\beta_1)$ follows
\begin{equation}
L(\beta_1) = 43.52-0.15\beta_1
\end{equation}
for 6 airlines (AA, DL, MQ, F9 and AS) except HA which has the lowest rate of delay in the 14 airlines so it is not in line of other airlines. For Model 2 with the results of ETSPL, the negatively correlated linear function (Fig.5(b))
$L(\beta_1)$ follows
\begin{equation}
L(\beta_1) = 38.83 - 1.32 \beta_1
\end{equation}
for 8 airlines (B6, UA, US, WN, EV, FL and OO) except VX which has the lowest $\beta_1(0.53)$ and the smallest number of flights except HA hence an ignorable propagation-delay effect so that its CCDF can decay more exponentially than that of any other airlines. From both Fig.2 and Fig.4, we reckon that the airline with smaller $\beta_1$ is much more effective to reduce the delay from last delay to current ones, i.e., it has a larger ability to absorb DP, which leads to the number of flights with small delay account for a larger proportion in all the delayed flights (See Fig.6).
 %Reversely speaking, this result can be understood by retrieving that the smaller the $\beta_2$($\sim \beta_1$) is, the steeper the $q(\it{l})$ decay in the same range $\it{l}$, the more effective the airline dealing with DP, hence the greater $L$ of the average absorption time of them.
 For examples behind, from Table 1 in $\bf{SI}$, among 6 airlines characterized by Model 1, F9 and AS behave less effective in the absorption of last delays with their larger $\beta_1$ compared with others.

\begin{figure*}[ht]
\centering
\includegraphics[width=0.9\linewidth]{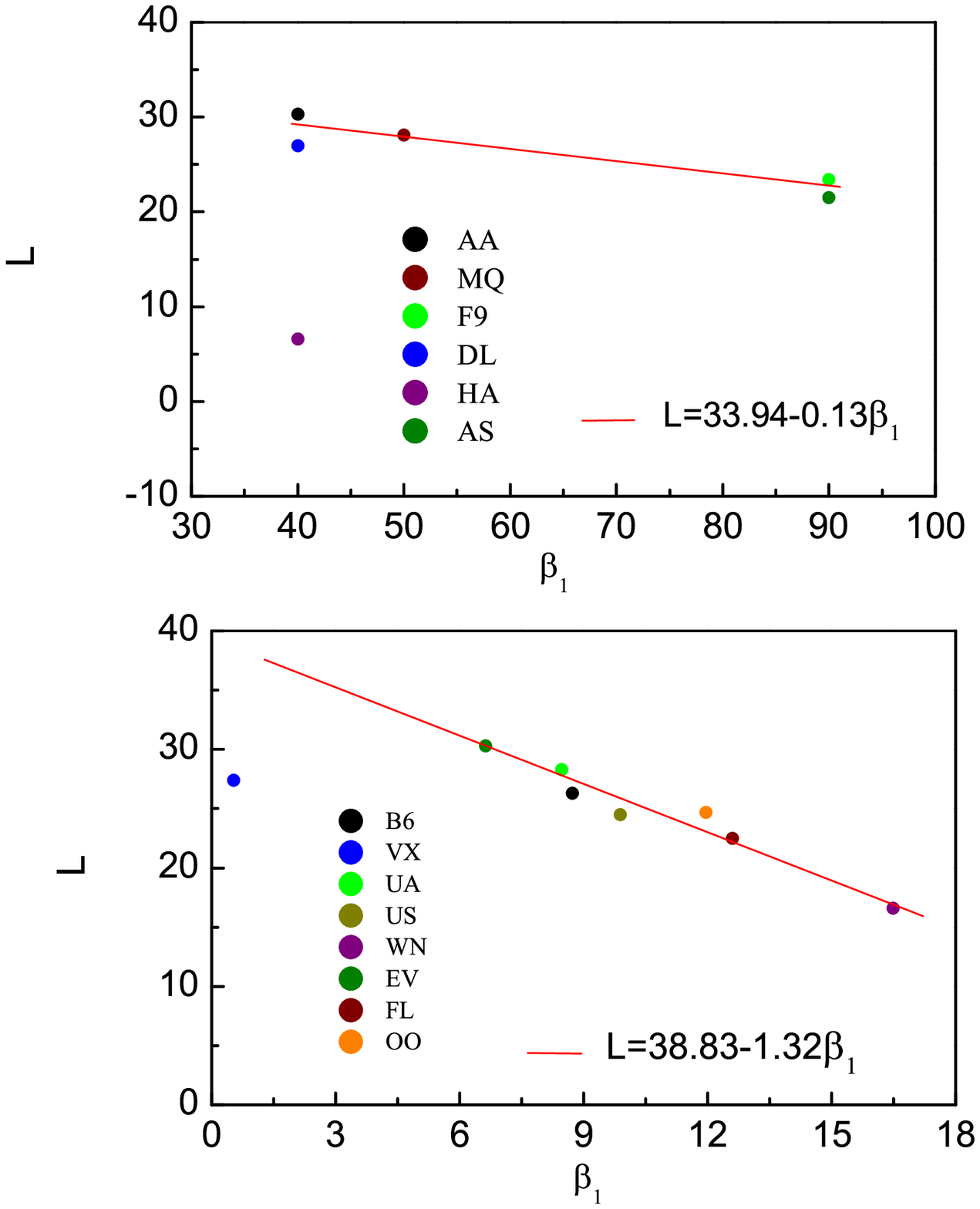}
\caption{(a)The aviation transportation meaning of shift parameter $\beta_1$ in Model 1 with the mechanism of DP. Average absorption time $L$ to DP is a negatively correlated linear function of $\beta_1$ except airline HA. $L = 39.4, 36.0, 12.0, 36.1, 32.3$ and $28.4$ for airlines AA, DL, HA, MQ, F9 and AS, respectively.(b)The aviation transportation meaning of shift parameter $\beta_1$ in Model 2 with the modified mechanism of DP. Average absorption time $L$ to DP is a negatively correlated linear function of $\beta_1$ except airline VX. $L = 27.4, 30.3, 28.3, 26.3, 24.5, 24.7, 22.5$ and $16.6$ for airlines VX, EV, UA, B6, US, OO, FL and WN, respectively.}
\label{fig:7}
\end{figure*}

%\begin{figure}[ht]
%\centering
%\includegraphics[width=1.1\linewidth]{Fig8}
%\caption{(b)The aviation transportation meaning of shift parameter $\beta_1$ in Model 2 with the modified mechanism of DP. Average absorption time $L$ to DP is a negatively correlated linear function of $\beta_1$ except airline VX. $L = 27.4, 30.3, 28.3, 26.3, 24.5, 24.7, 22.5$ and $16.6$ for airlines VX, EV, UA, B6, US, OO, FL and WN, respectively.}
%\label{fig:8}
%\end{figure}

\begin{figure*}[ht]
\centering
\includegraphics[width=0.9\linewidth]{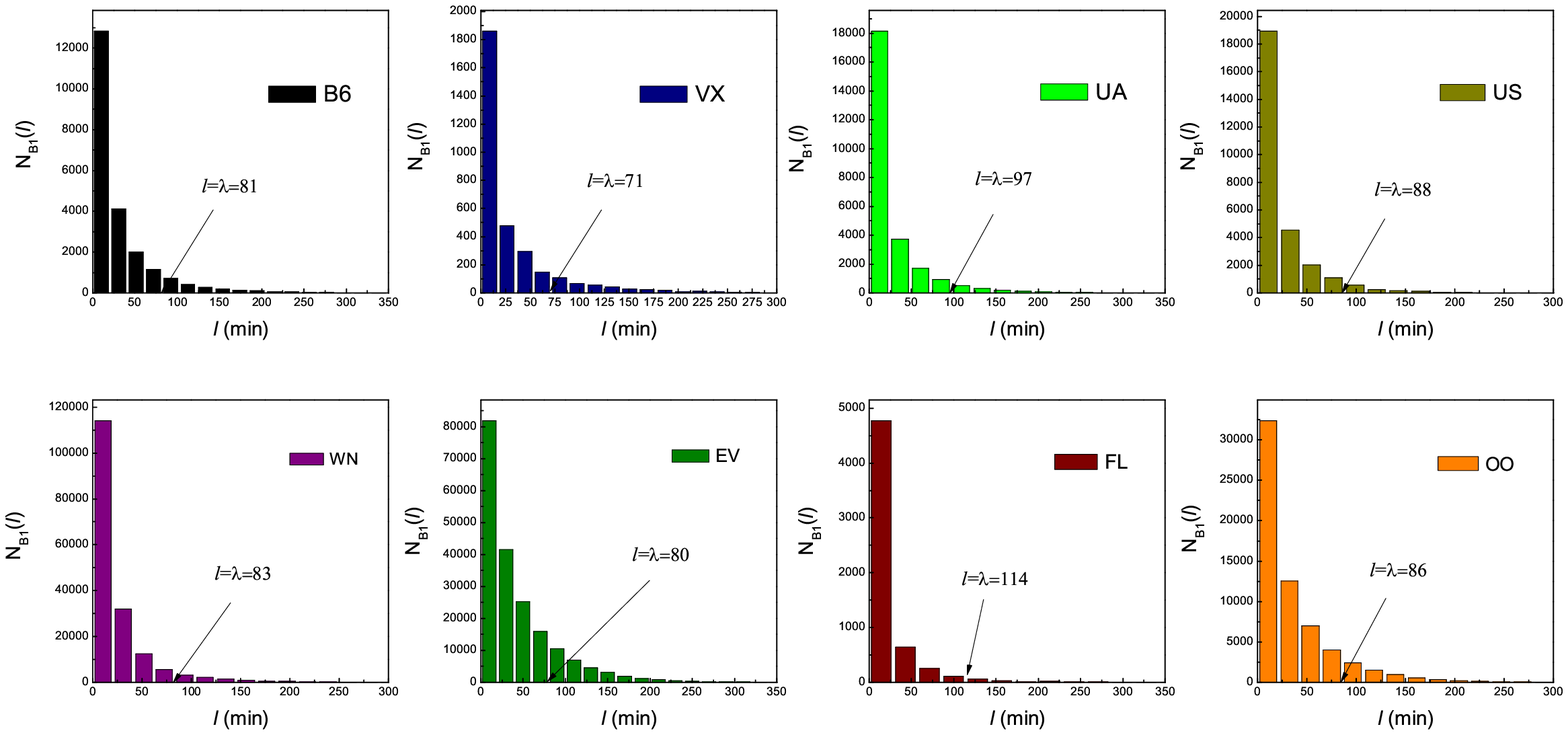}
\caption{Number distributions $N_{B1}(\it{l})$ of flights with DP in 8 airlines are biased to the side with
$\it{l} < \lambda$. Histograms are assigned with equal width $\Delta \it{l} = \frac{1}{4} \lambda$. Arrows indicate the positions of critical value $\lambda$ separating the dominant range PF from NPF for DD of airlines B6, VX, UA, US, WN, EV, FL and OO, respectively. The functions $N_{B1}(\it{l})$ means the flight number caused by PF is based on direct counting of flights with larger delays than the current one, without the compensation interval $m$.}
\label{fig:9}
\end{figure*}

In comparison with CCDFs of 6 airlines fit by SPLs, integrated CCDFs of 8 airlines drop down steeper as increased delay $\it{l}$, and they have smaller $\beta_1$, which means that the factor $(\it{l}+\beta_1)^{-r}$ behaves roughly as $\it{l}^{-r}$ for the cases $\it{l} > \lambda$ because $\lambda > \beta_1$ (see Table 2 in $\bf SI$) is always satisfied. Actually, since smaller $\beta_1$ means larger absorption $L(\beta_1)$, therefore the probability to have propagation-caused delay from larger $\it{l}$ decreases with $\it{l}$ quickly. While the chance for NPF to act on delay increases.
% which induces an exponential-like function of probability distribution $P(\it{l})$, forming a clear but trivial result.
In this category of delay cases, $\lambda$ acts more importantly than $\beta_1$. The smaller $\lambda$ an airline has, the smaller interval for it to have a propagation- induced delay with longer delay than that of current one. Among 8 airlines, VX behaves as the typical case since it has both small $\beta_1$ and $\lambda$. While FL has larger probabilities with the effects of DP than others since it has both larger $r$ as the power in SPL and larger parameter $\lambda$ emphasize DP dominating tendency. However, such a picture does not mean smaller fraction of flights with DP since the actual distribution for a flight to delay is on average biased to the side with $\it{l} < \lambda$, even with the fraction as high as 90 percent, which is shown in Fig.6 for 8 airlines covered by Model 2. The delayed number $N_{B1}(\it{l})$  means the flight number caused by PF is based on direct counting of flights with larger last delays than the current one, without the compensation interval $m$ in Model 2, for the comparison among airlines with equal conditions. while the width of the histograms is set to be $\frac{1}{4}\lambda$ for clear look in Fig.6, with the arrows indicating the positions of critical delay $\lambda$ separating the PF dominating range from that of NPFs.

\section*{Conclusions}

     In the present work, two new mechanisms are presented to model flight delays based on the big data from BTS of the United States. Two types of CCDFs describing delays are obtained from  two models with analytical derivation, they fit empirical data well. Fourteen American airlines can be sorted into two categories. One is more dominated by the propagation of delays, and its CCDF follows SPL. The other is more dominated by PF under a critical delay $\lambda$, while it is more dominated by NPF above it, and its CCDF follows ETSPL.

      Starting from the mechanism of delay propagation, we convert the effect from NPF and cancellations into it, not only find the key parameter $\beta_1$ for the mechanism of propagation, but also find the compensation interval $m$ describing the overlapping effects with NPF and smaller last delays than current ones. Moreover, key parameter $\lambda$ separating the effect of PF from NPF also emerges out. Three quantities demonstrate their practical meaning in aviation, which can serve as new metrics to measure the quality of management and operation of airlines. In addition, by accounting the current delays as the transmission result of all previous delays incurred by all kinds of factors, we carry the spirit of mean-field approach in temporal regime. The approach that converting other factors into an equivalent propagation one before extracting its quantitative aviation meaning verifies its validity in dealing with temporal big data, which is hopefully applicable to many topics in wide other fields.

%\section*{Methods}

\section*{Acknowledgements}

This research was supported by the National Natural Science Foundation of China (Grant Nos. 61304190, 11175086), the Fundamental Research Funds for the Central Universities (Grant No. NJ20150030).

\section*{Author contributions statement}

Y. J. W., C. P. Z. and H. E. S. designed the study; Y. J. W and F. W collected and rectified data, Y.K. C derived formulas, C.P.Z, Y. K. C, Y. J. W., M. H. H. and B.B analyzed results, C.P. Z, Y. J. W, Y. K. C, B. B. and H. E. S. wrote the paper.

\dag Y. J. W and Y. K. C contributed equally to the paper.
\section*{Additional information}
oldpigamn1234@126.com;

ywang@nuaa.edu.cn;

hes@bu.edu

%\textbf{Supplementary information}


\begin{thebibliography}{99}


\bibitem{Fleurquin}
Fleurquin P, Ramasco J J, Eguiluz V M. Characterization of Delay Propagation in the US Air-Transportation Network. Transportation Journal, 2014, 53(3): 330-344.


\bibitem{Kafle16}
Kafle N, Zou B. Modeling flight delay propagation: A new analytical-econometric approach. Transportation Research Part B: Methodological, 2016, 93: 520-542.


\bibitem{Ahmad08}
AhmadBeygi S, Cohn A, Guan Y, et al. Analysis of the potential for delay propagation in passenger airline networks. Journal of air transport management, 2008, 14(5): 221-236.

\bibitem{Ahmad10}
Ahmadbeygi S, Cohn A, Lapp M. Decreasing airline delay propagation by re-allocating scheduled slack. IIE transactions, 2010, 42(7): 478-489.

\bibitem{Wang03}
Wang P T R, Schaefer L A, Wojcik L A. Flight connections and their impacts on delay propagation//Digital Avionics Systems Conference, 2003. DASC'03. The 22nd. IEEE, 2003, 1: 5. B. 4-5.1-9 vol. 1.

\bibitem{Beatty}
Beatty R, Hsu R, Berry L, et al. Preliminary evaluation of flight delay propagation through an airline schedule. Air Traffic Control Quarterly 7(4), 259-270 (1999).

\bibitem{Lan}
Lan S, Clarke J P, Barnhart C. Planning for robust airline operations: Optimizing aircraft routings and flight departure times to minimize passenger disruptions[J]. Transportation science, 2006, 40(1): 15-28.


\bibitem{Campanelli}
Campanelli B, Fleurquin P, Arranz A, et al. Comparing the modeling of delay propagation in the US and European air traffic networks. Journal of Air Transport Management, 2016.


\bibitem{Fleurquin13}
Fleurquin P, Ramasco J J, Eguiluz V M. Systemic delay propagation in the US airport network. Scientific reports, 2013, 3.

 \bibitem{Liu}
 Liu Y, Hansen M, and Zou B. Aircraft gauge differences between the US and Europe and their operational implications. Journal of Air Transport Management, vol. 29, pp. 1-10, 6// 2013.



\bibitem{Fricke}
Fricke H, Schultz M. Delay impacts onto turnaround performance//ATM Seminar. 2009.

\bibitem{Abdelghany}
Abdelghany K F, Shah S S, Raina S, et al. A model for projecting flight delays during irregular operation conditions. Journal of Air Transport Management, 2004, 10(6): 385-394.

\bibitem{Pyrgiotis}
Pyrgiotis N, Malone K M, Odoni A. Modelling delay propagation within an airport network. Transportation Research Part C: Emerging Technologies, 2013, 27: 60-75.


\bibitem{Tu}
Tu Y, Ball M O, Jank W S. Estimating flight departure delay distributions-a statistical approach with long-term trend and short-term pattern. Journal of the American Statistical Association, 2008, 103(481): 112-125.


\bibitem{Wong}
Wong J T, Tsai S C. A survival model for flight delay propagation[J]. Journal of Air Transport Management, 2012, 23: 5-11.


\bibitem{BTS}
http://www.bts.gov


\bibitem{Baumgarten}
Baumgarten P, Malina R, Lange A. The impact of hubbing concentration on flight delays within airline networks: An empirical analysis of the US domestic market. Transportation Research Part E: Logistics and Transportation Review, 2014, 66: 103-114.

%\bibitem{Barabasi}
%Barabasi A L, Albert R. Emergence of scaling in random networks[J]. science, 1999, 286(5439): 509-512.

\bibitem{Albert}
Albert R, Barabasi A L. Statistical mechanics of complex networks. Reviews of modern physics, 2002, 74(1): 47.

\bibitem{Zhou}
Zhou T, Yan G, Wang B H. Maximal planar networks with large clustering coefficient and power-law degree distribution. Physical Review E, 2005, 71(4): 046141.

\bibitem{Han}
Han X P, Zhou T, Wang B H. Modeling human dynamics with adaptive interest. New Journal of Physics, 2008, 10(7): 073010.

\bibitem{Zhou08}
Zhou T, Kiet H A T, Kim B J, et al. Role of activity in human dynamics. EPL (Europhysics Letters), 2008, 82(2): 28002.

%\bibitem{Barabasi09}
%Barabasi A L. Scale-free networks: a decade and beyond[J]. Science, 2009, 325(5939): 412-413.

%\bibitem{Oliveira}
%Oliveira J G, Barabasi A L. Human dynamics: Darwin and Einstein correspondence patterns[J]. Nature, 2005, 437(7063): 1251-1251.

\bibitem{Chang}
Chang H, Su B B, Zhou Y P, He D R, Assortativity and act degree distribution of some collaboration networks. Physica A: Statistical Mechanics and its Applications, 2007, 383(2): 687-702.


\bibitem{Fu}
Fu C H, Zhang Z P, Chang H, et al. A kind of collaboration $-$competition networks. Physica A: Statistical Mechanics and its Applications, 2008, 387(5): 1411-1420.


\bibitem{Wang}
Wang Y L, Zhou T, Shi J J, et al. Empirical analysis of dependence between stations in Chinese railway network. Physica A: Statistical Mechanics and its Applications, 2009, 388(14): 2949-2955.

\bibitem{Gonzalez}
Gonzalez M C, Hidalgo C A, Barabasi A L. Understanding individual human mobility patterns. Nature, 2008, 453(7196): 779-782.

%\bibitem{Newman}
%Newman M E J. Scientific collaboration networks. I. Network construction and fundamental results[J]. Physical review E, 2001, 64(1): 016131.


%\bibitem{Yan}
%Yan X Y, Han X P, Wang B H, et al. Diversity of individual mobility patterns and emergence of aggregated scaling laws[J]. Scientific reports, 2013, 3.

\bibitem{Kwak}
Kwak H, Lee C, Park H, et al. What is Twitter, a social network or a news media?//Proceedings of the 19th international conference on World wide web. ACM, 2010: 591-600.


\bibitem{absorption1}
Yablonsky G, Steckel R, Constales D, Farnan J, Lercel D, Patankar M, Flight delay performance at Hartsfield-Jackson Atlanta International Airport, J. Airline and Airport Management, 2014-4(1), 78-95.


\bibitem{absorption2}
Hunter C. G. and Weidner T., Performance and Benefits Modeling of Center Airspace ATM Improvements, AIAA GNC Conference, 1398, 1997-arc.aiaa.org.





\end{thebibliography}
\end{document}

\title{Supplementary Information}
\maketitle
\section*{Supplementary Figs}
\begin{figure*}[ht]
\centering
\includegraphics[width=0.8\linewidth]{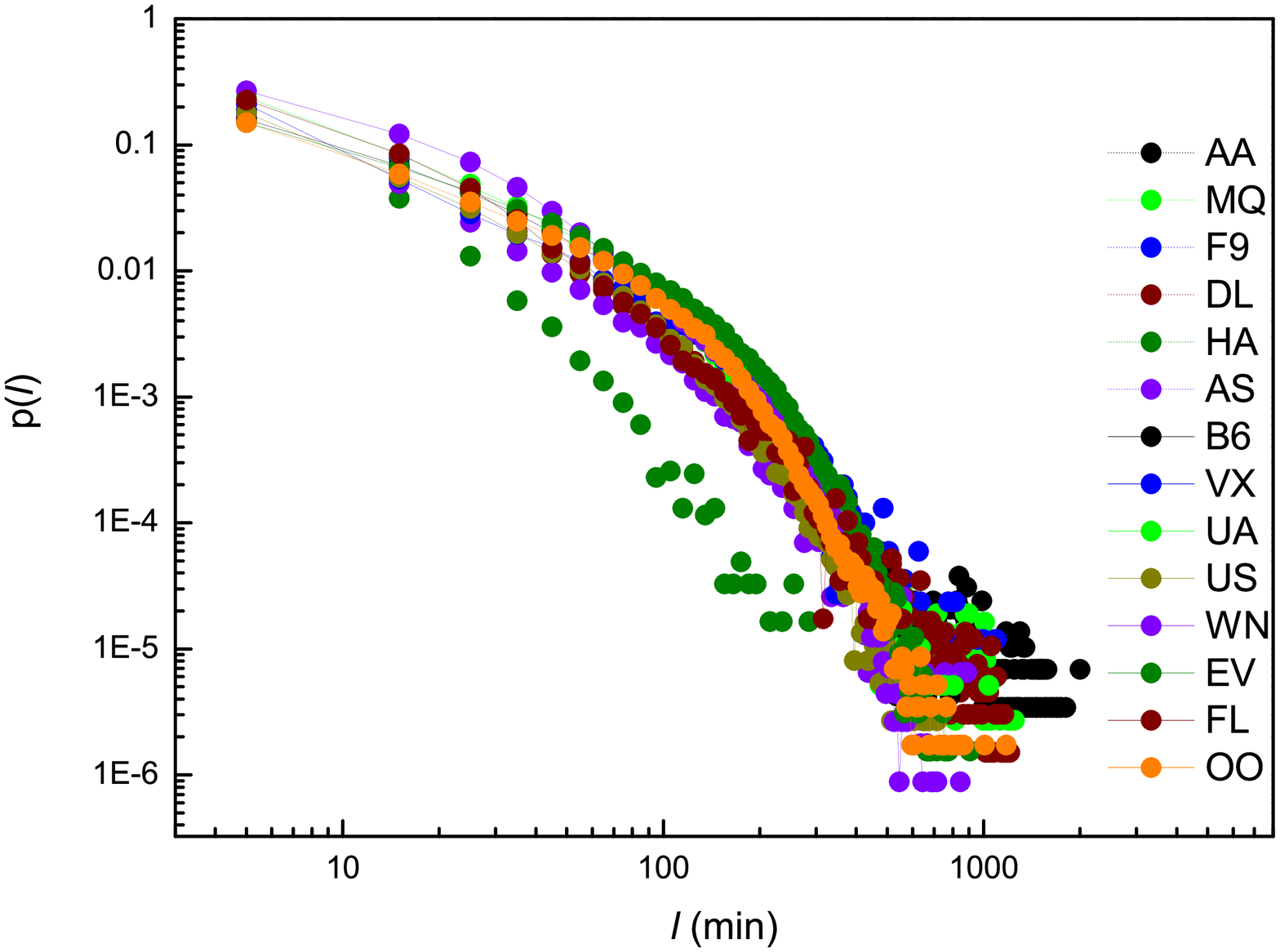}
\caption{Probability distribution functions $p(\it{l}) $ of departure delays $\it{l}$ for 14 airlines of United States. $\Delta \it{l} = 10 min$}
\label{fig:1}
\end{figure*}
\begin{figure*}[ht]
\centering
\includegraphics[width=0.8\linewidth]{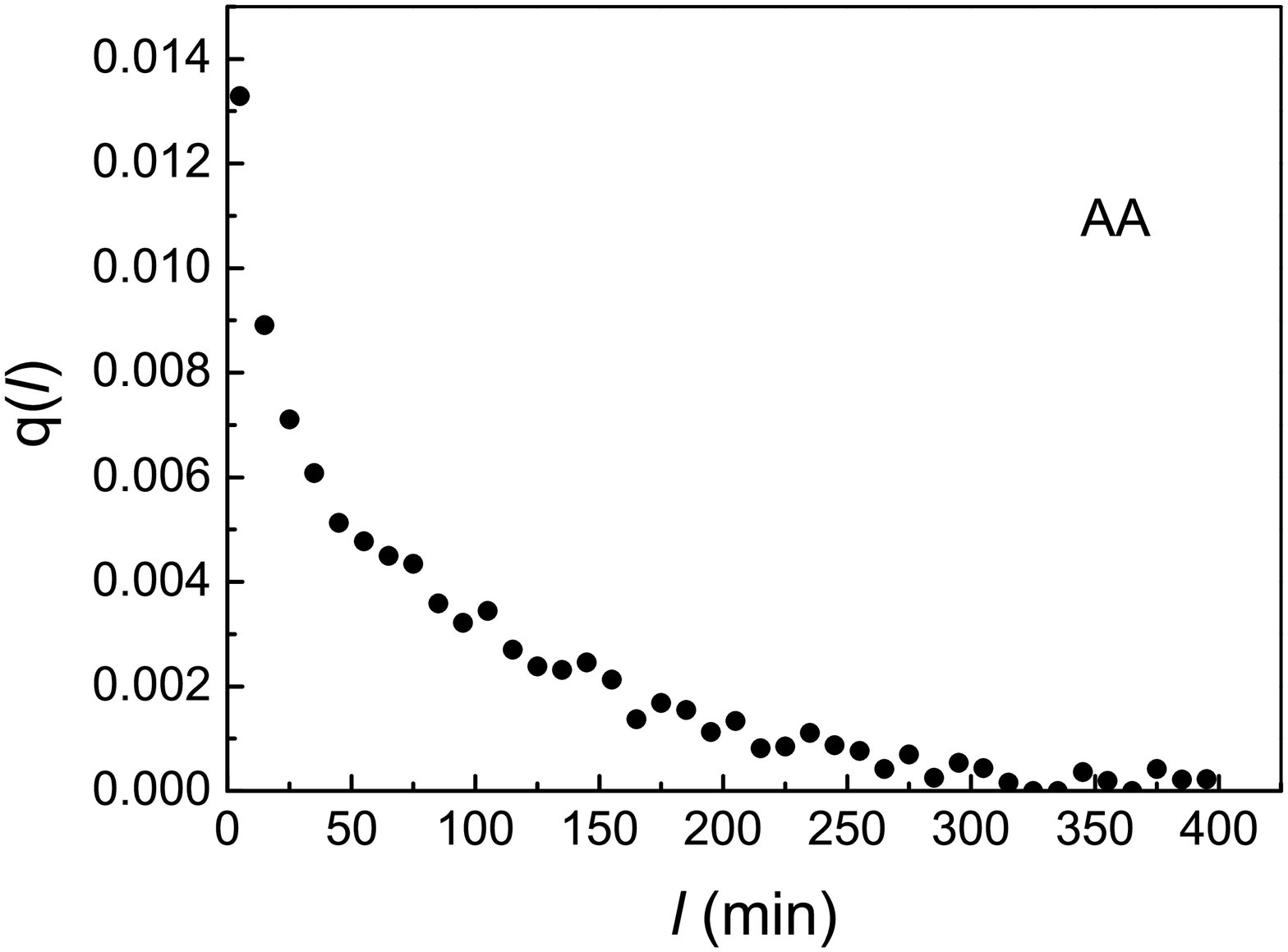}
\caption{Transmission probability per delay for the flights with the delay in the temporal range $(\it{l}, \it{l}+\Delta \it{l})$
caused by the propagation from the range $(\it{l}, \infty)$.
 Filled circles are obtained from the cumulated statistics on primitive
records of airline AA with $\Delta \it{l} = 10 min$.
}
\label{fig:2}
\end{figure*}

\newpage
\section*{Check the tentative function $q(\it{l})$ underlying SPL of CCDFs.}
 At the tails of CCDF curves with colored symbols from empirical data, obvious dropping down from linear fitting lines are always observed. From Fig.2 we see that the probability for a flight to be transmitted by a very long delay is very small, it only fluctuates a little bit above zero. So the behavior of actual flights do not follow SPL in the limit of large $\it{l}$. Actually, flights may be canceled when they delay too much. And the longer one delays, the larger probability for it to be canceled.  However, formula (8) is derived under a potential prerequisite, i.e., the mechanism of DP is valid for any delay, which is not true, especially for very large delays. In other words, assumption (1) becomes ineffective in such tails.

The effectiveness of SPL shown in formula (8) seems to rely on the validation of the tentative function $q(\it{l})$. It is checked by the comparison with the results from  empirical data. The functions $q(\it{l})$ are integrated for smooth look in Fig.3 and Fig.4. The filled circles in  black, blue, green, dark yellow, purple,
olive, wine red and orange represent integrated $q(\it{l})$ of airlines AA, MQ, F9, DL, HA and AS based on the definition (formula (2)), each with shift parameters $\beta_1 = 35, 40, 90, 30, 30$ and $80$, respectively. The red lines represent the tentative function in the integrated form of fractional function $q(\it{l}) = \frac{\alpha}{\it{l}+\beta_2}$ from formula (3). And the integrated form $F(\it{l})$ with variable upper bound reads: $F(\it{l}) = \int_{\it{l}}^{\infty}\alpha (ln(\it{l}+\beta_2 ) - ln \beta_2) $. They fit linear parts of filled circles representing empirical data of these 6 airlines well. Due to the same reason in fitting curves in Fig.3, $F(\it{l})$ deviates from increasing red lines and gets saturated when $ \it{l}$ becomes large enough.

The parameters used in the integrated CCDFs and the function $F(\it{l})$,
%i.e., integrated from of the tentative function $q(\it{l}$
are collected in Table 1 which shows obvious inconsistence between corresponding columns $\beta_1$ and $\beta_2$ for airlines AA, MQ, DL and AS, respectively. $n_B(\it{l},\it{l}+\Delta \it{l})$ from formula (1) to (3) is defined as the number of current delayed flights per delay caused by last delayed flights and this number is actually not discernable in the practical data, which has been pointed out \cite{Abdelghany} before the present work. So, in our simulation, we have to actually substitute it by the number of last flights with larger delay than the current one. This manipulation is carried out in Fig.3 and Fig.4 for plotting filled circles from real data, which is not always true because such substitution results in smaller quantities of $n_B(\it{l},\it{l}+\Delta \it{l})$ than the existing ones in fact. However, filled circles in Fig.1 and Figf.2 in the main text are plotted by the data directly from the primary records. And by fitting red lines to them the parameters $\alpha_0$ and $\beta_1$ are obtained. Neither the data symbols nor the fitting line in Fig.1 and Fig.2 in the main text rely on the substitution of the parameters $\alpha_1$ and $\beta_2$ into them from Fig.3 and Fig.4. Actually integrated CCDFs in Fig.1 and Fig.2 have included the effect from all kinds of delay factor calibrated into equivalent DP measure in formula (5). Therefore, such two parameters $\alpha_0$ and $\beta_1$ serve as the baselines for the comparison.  Difference between $\beta_2$ and $\beta_1$ actually reveals the degree of effectiveness of that manipulation in the simulation and the assumption of all-DF, which is also used to understand the earlier deviation of lines $F(\it{l})$ from the colored data symbols in the variable $\it{l}+\beta_2$ in the panels of Fig.3 and Fig.4.

\newpage
\section*{Tables for Model 1 and Model 2}
\begin{table}
\caption{ Comparison of the parameters in DDCFs for model 1 \\
with corresponding ones in the transmission function $q(\it{l})$ .}
\label{tab:datasets}
\begin{tabular}{llllll}
\hline
    airlines  & $ c_0$  & $\beta_1$ & $\alpha_0$     $|$& $\alpha$ & $\beta_2$ \\
\hline
\hline
  AA   & $7.77 \times 10^2$  &    40  & 2.03 $|$& 0.43 &  35  \\
\hline
  MQ    & $3.74\times 10^3$ &   50   & 2.84 $|$& 0.49 &  40   \\
\hline
  F9    & $5.37\times 10^6$ &   90  & 3.59 $|$& 0.59 &  90   \\
\hline
  DL   &  $4.67\times 10^3$ &  40   &  2.51 $|$& 0.37&  30   \\
\hline
  HA   &  $9.48\times 10^4$ &  30   &  3.78 $|$& 0.70&  30   \\
\hline
  AS   &  $7.79\times 10^6$ &  90   &  3.86 $|$& 0.71&  80   \\
\hline
\hline
\end{tabular}
\end{table}
\begin{table}
\caption{ Comparison of the parameters in DDCFs for model 2 \\
with corresponding ones in the transmission function $q(\it{l})$ .}
\label{tab:datasets}
\begin{tabular}{llllllll}
\hline
airlines      $|$& $c_2$  & $\beta_1$ & $    \lambda$ &     $r$     $|$&$\alpha$ & $\beta_2$ & $m$ \\
\hline
\hline
  B 6          $|$&  1.18  & 8.73   &81.97  &0.50 $|$& 0.52  &8.70   &  15\\
\hline
  VX          $|$& 0.35  &  0.53   & 70.92 &0.22 $|$& 0.33 & 0.52  &  15\\
\hline
  UA          $|$& 1.99  &  8.47    &97.07 &0.63 $|$& 0.41  &8.45   &  8 \\
\hline
  US          $|$& 2.13  &  9.90    & 87.72 &0.79 $|$& 0.48  &9.25  &  10\\
\hline
WN            $|$& 8.73  & 16.48    & 82.64 &0.94 $|$& 0.74 & 16.77 & 6 \\
\hline
  EV          $|$& 0.79  &  6.62    & 70.37 &0.33 $|$& 0.53 & 6.86  & 21\\
\hline
  FL          $|$& 6.95  & 12.61    & 113.90&1.08 $|$& 0.55 & 16.39 & 7 \\
\hline
OO            $|$& 1.54  & 11.64    & 86.21 &0.59 $|$& 0.75 & 11.94 & 24\\
\hline
\hline

\end{tabular}
\end{table}